\documentclass[]{aa}
\usepackage{natbib}
\usepackage{psfig}
\usepackage{graphicx}
\usepackage{txfonts}
\def\HH{\mbox{H$_2$}}

\def\nH2{{\rm n}({\rm H}_2)}
\def\NH2{{\rm N}({\rm H}_2)}
\def\pccc{~{\rm cm}^{-3}} 
\def\pcc {~{\rm cm}^{-2}}

\def\Trs {\mbox{${\rm T}_{\rm r}^*$}}
\def\Tsub#1 {\mbox{${\rm T}_{\rm #1}$}}
\def\TK  {\Tsub K }

\def\Tmb {\Tsub mb }

\def\arcsec{\mbox{$^{\prime\prime}$}} \def\arcmin{\mbox{$^{\prime}$}}

\def\degr{\mbox{$^{\rm o}$}}

\def\p{\mbox{$^+$}}
\def\cotw {\mbox{$^{12}$CO}}

\def\WCO{\mbox{W$_{\rm CO}$}}

\def\h13cop{\mbox{{H$^{13}$CO\p}}}

\def\c3h2{\mbox{C$_3$H$_2$}}
 
 \def\R0{R$_0$}

\def\ddeg{{}^\circ\kern-.1em}  
\def\Msun{\mbox{M$_{\rm sun}$}}  

\def\kms{\mbox{km\,s$^{-1}$}}

\def\E#1 {$10^{#1}$}
\def\E#1 {E{#1}}
\def\P#1,{$\nH2\TK~=~#1\times~10^4\pccc$~K}
\def\ec#1,#2,#3,{#1\,(#2)\E{#3}}

\def\H3{\mbox{H$_3$}}

\def\RH2{\mbox {R$_{\rm G}$}}
\def\fH2{\mbox {f$_{\HH}$}}
\def\FH2{\mbox {F$_{\HH}$}}

\title{CO J=1-0 observations of molecular gas interacting with galactic supernova remnants 
  G5.4-1.2, G5.55+0.32 and G5.71-0.08 \thanks{Based on observations obtained with the ARO Kitt Peak 12m 
   telescope.}}
\author{H. S. Liszt\inst{} }
\institute{National Radio Astronomy Observatory,
           520 Edgemont Road,
           Charlottesville, VA,
           USA 22903-2475}

\begin{document}
\date{received \today}
\offprints{H. S. Liszt}
\mail{hliszt@nrao.edu}
%

\abstract
   {The field just West of the galactic supernova remnant W28 
  (l=6.4\degr, b=-0.2\degr) harbors 3 of 5 newly-discovered 1720 OH maser spots
  and  two recently-discovered  candidate supernova candidates (one of which
  is a $\gamma$-ray source), as well as several compact and classical HII regions.}
    {To show the interaction of radio supernova remnants with ambient
   molecular gas in sky field just West of W28.}
   {We analyze a datacube of CO J=1-0 emission having 
  1\arcmin\ and 1 \kms\ resolution, made with on-the-fly mapping over 
  the region $5\degr \le l \le 6\degr, -1\degr \le b \le 0.5\degr$}
  {Extended and often very bright CO emission was detected at the velocities 
  of the 1720 MHz OH masers and around the supernova remnant 
 G5.55+0.32 which lacks a maser.   
  A new bipolar outflow which is marginally resolved at 1\arcmin\ 
   resolution and strong in CO (12K) was detected at the periphery of 
  G5.55+0.32, coincident with an MSX source;  there is also a bright
  rim of CO just beyond the periphery of the radio remnant.
  The OH maser near G5.71-0.08 lies on a shell of strongly-emitting
 molecular gas (up to 20K) .   
  At the -21 \kms\ velocity of G5.4-1.2, CO covers much of the 
   field but is weak (3 K) and undisturbed near the remnant.
  The extended molecular gas around the compact H II region and 
  outflow in G5.89-0.39 (W28A2) is shown for the first time.}
 {}

\titlerunning{When things collide}

\maketitle

\section {Introduction.}

Luminous early-type stars exert strong influence over the galactic interstellar
medium (ISM) at all phases of their evolution, whether as outflows from
the protostar, winds and sources of ionizing photons from the 
main-sequence object,
or as radiative and mechanical drivers in the end-stage supernova
remnant (SNR).   In a few cases, the end-stage interaction of an SNR 
with molecular gas in the ambient ISM is apparent and these special
cases have been studied with great interest;  particular
examples include IC443 and W44 \citep{Den79,Den83}, 
W28 \citep{Woo81,Den83,AriTat+99,AhaAkh+08}, Vela \citep{MorYam+01}
and others noted by \cite{AriTat+99}, \cite{ReyMan00} and \cite{MorTam+05}

\begin{figure*}
\psfig{figure=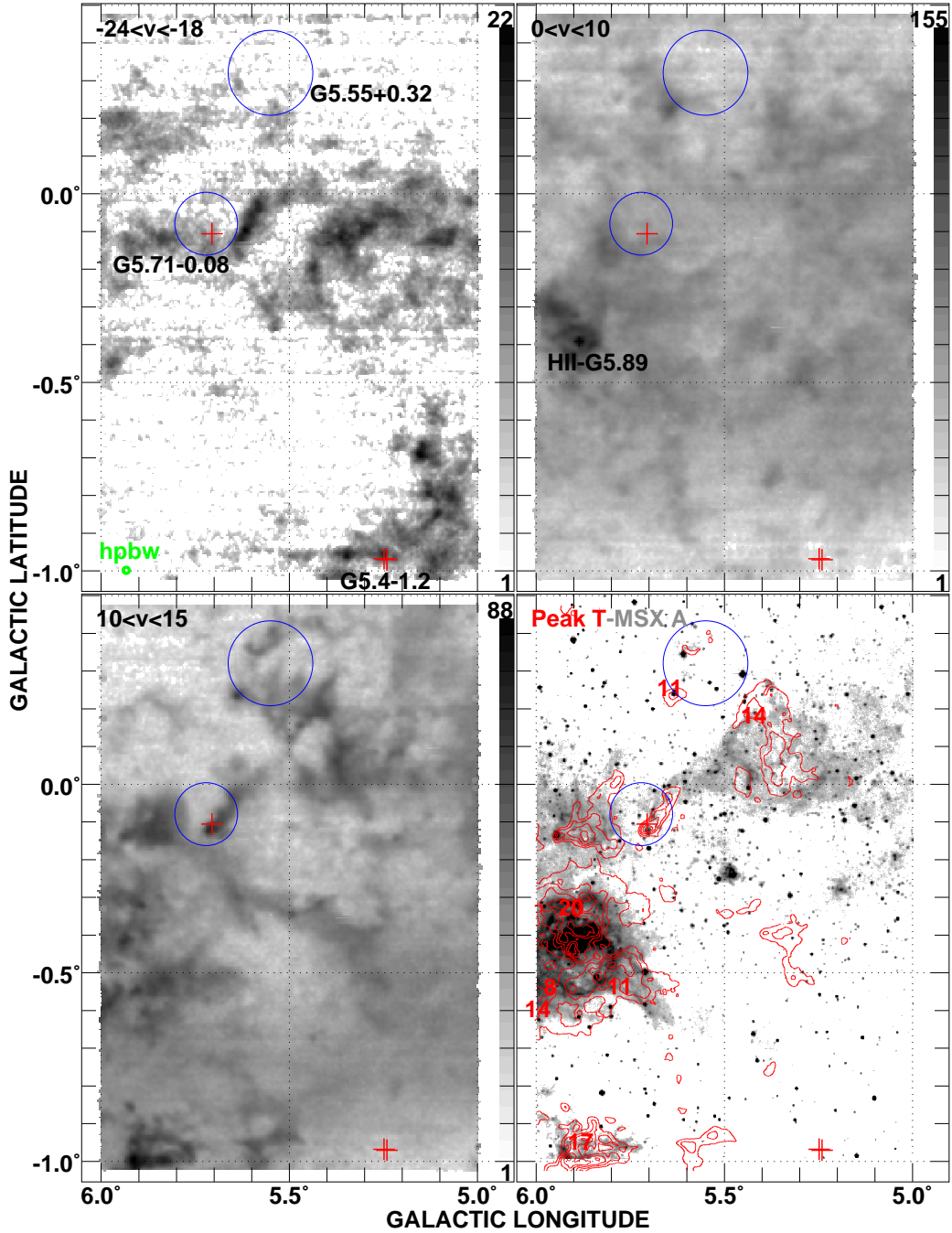,height=18cm}
\caption[]{CO J=1-0 and MSX 8.3$\mu$IR emission in the field around 
l=5.5\degr.
In the panel at lower right, a map of the peak CO temperature 
at $|v| < 26$ \kms\ is shown as (red) contours of temperature in 
Kelvins at levels 8,11,14,17,20 K superposed on an image of MSX
Band A emission around 8.3$\mu$.  In the other panels, the grey
scale represents CO emission (\WCO) integrated over the velocity
ranges indicated in each panel and the bar scales show the gradation
of intensity and the peak integrated line brightness \WCO .  The 1\arcmin\ 
ARO 12m telescope HPBW is
shown in the upper left panel.  Crosses (red) mark the positions
of three newly-detected 1720 MHz OH maser spots \citep{HewYus09}
and the outlines of two new SNR ``candidates'' discovered by
 \cite{BroGel+06} are sketched as (blue) circles.}.
\end{figure*}

Study of the SNR-ISM interaction has been hindered by the relative 
paucity of identified SNR compared with accepted rates of star death 
and by the absence of large-scale, high-resolution surveys of the 
OH 1720 MHz and CO emission which are typically used to trace the
interaction.  However, new radiocontinuum surveys of the galactic plane 
\citep{HelBec+06,BroGel+06} have recently identified a host of (candidate)
SNR in the inner galaxy and surveys searching for signposts of 
interaction in the 1720 MHz transition of OH have been conducted by 
\cite{HewYus09}.

The sky just to the galactic West of W28 (G6.4-0.2) is particularly 
interesting in this context because two new candidate SNR were noted
by \cite{BroGel+06} and searches of these objects returned 3 of only 5
newly-detected 1720 MHz OH maser spots (out of 35 fields searched)
in the work of \cite{HewYus09}.  Moreover, the field is a copious HESS
source of $\gamma$-rays coincident with several  SNR and the 
nearby molecular gas mapped in CO with 4\arcmin\ resolution at NANTEN 
\citep{AhaAkh+08}.  Also in the field is the very bright ultra-compact
HII region G5.89-0.39 (W28-A2) with a well-studied molecular
outflow \citep{KlaPlu+06,StaGos+07,WatChu+07,HunBro+08}.

By chance, the field west and south of W28 was recently mapped at 
comparatively high resolution (1\arcmin) in CO J=1-0 emission
for other purposes, namely to show the intake and shredding of 
gas into the galactic dust lane running along the large-scale
galactic bar \citep{Lis06Shred}. As a result, only the high-velocity
portion of the emission and the more westerly portion of the 
field were discussed.  However, the full datacube contains a 
wealth of information on the molecular gas, HII regions and SNR
in the more nearby regions of the Galaxy, and this is discussed here 
for the first time.

The organization of this paper is as follows.  Section 2 summarizes
the CO datacube.  Section 3 shows the sky in this region and Section
4 is a brief summary.

\section{The data and nominal conversion to \HH\ mass}

The CO datacube which forms the basis of this work was presented
originally by \cite{Lis06Shred} and can now be obtained from the
SIMBAD site.  [NB: SIMBAD tells me that they will host the cube if the 
paper is published, for now the cube maybe downloaded from 
ftp://tinyurl.com/ftphsl/broad5co.flb]  

The region $5\degr \le l \le 6\degr, -1\degr \le b \le 0.5\degr$
was mapped at the ARO 12m telescope on several successive days in 
2002 May in on-the-fly mode, resulting in a map which is fully 
sampled in space at 1\arcmin\ resolution on a 15\arcsec\ pixel 
grid, with channel spacing 390.6 kHz or 1.016 \kms\ and slightly 
lower resolution, 1.2 \kms. The single-channel rms noise level is 
typically 0.25 K.  The brightness scale of the ARO 12m telescope
is on the \Trs\ scale and \Tmb $\approx$ \Trs/0.85.  All CO brightnesses
quoted here are on the native \Trs\ scale and all velocities are
with respect to the Local Standard of Rest.

Nominal \HH\ masses associated with the CO emission are calculated
assuming a typical conversion factor N(\HH) $= 2\times10^{20}\pcc$ \WCO\
where \WCO\ is an integrated brightness of CO in units of K-\kms.

\section{Disk and bar gas near the galactic plane at 5\degr\ $< l < 6$\degr }

The sky is shown in the four panels of Fig. 1 and illustrative profiles
at four pixels are shown in Fig. 2.  The galactic coordinates of
the pixels are shown in each of the panels of Fig. 2.

In Fig. 1 the positions and approximate extent of the SNR G5.55+0.32 
and G5.71-0.08 are illustrated by (blue) circles; the presence of 
SNR G5.4-1.2 is noted at the bottom of the upper left panel but the
body of the remnant lies outside these images.  W28 lies 0.4\degr\ East 
of the left edge of these maps.  The new 
1720 MHz OH maser spots discovered by \cite{HewYus09} are shown 
as (red) crosses and the 1\arcmin\ telescope beam (hpbw) is shown in the 
upper left panel.  The HII region W28A2 = G5.89-0.39 is noted in 
the upper right panel.  Another, much larger and brighter H II 
region and strong MSX source at l=6\degr, b=-1.2\degr\ is not
apparent in Fig. 1 but it extends into the lower left corner of
the mapped region and is responsible for the very bright CO which is found 
there (see the lower panels).

Except at lower right, the panels show CO emission integrated over
velocity ranges chosen to contain a 1720 OH maser or other aspect; 
the  range is indicated at upper left in each panel and the integrated
brightness levels (\WCO) are indicated by the bar scale.  The panel at 
lower right shows (red) contours of the peak CO brightness over the 
wider range -26 \kms\ $<$ v $<$ 26 \kms\ overlaid on an image of MSX 
\citep{PriEga+01} Band A 8.3$\mu$ emission.

\begin{figure}
\psfig{figure=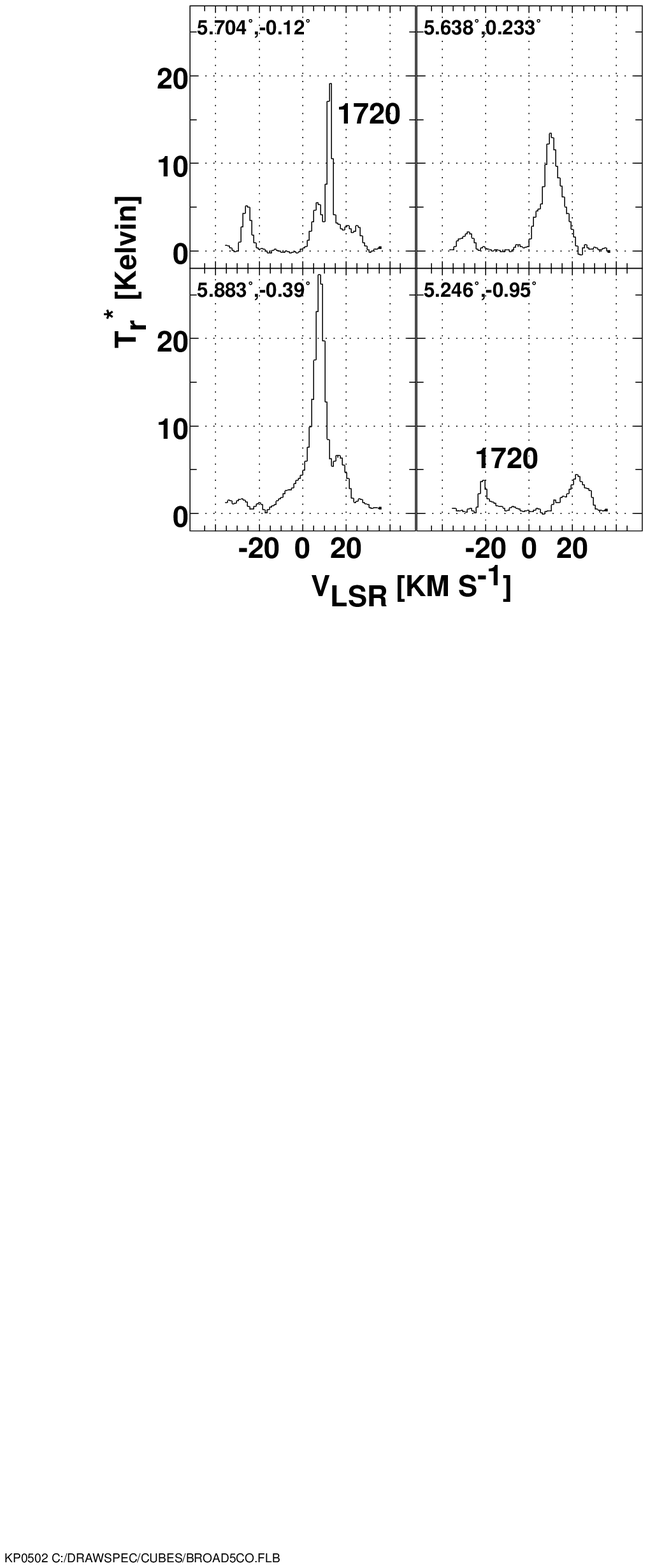,height=10cm}
\caption[]{ARO \cotw] profiles at 4 pixels in Fig. 1.  Galactic
coordinates are indicated in each panel.  The channel spacing is
is 1.016 \kms\ (resolution 1.2 \kms) and the spatial resolution is 
1\arcmin.}
\end{figure}

\subsection{G5.4-1.2}

Two of the new OH maser spots found by \cite{HewYus09} occur in 
the lower right portion of our maps but the CO line in the region 
is entirely undistinguished, see Fig. 2 at lower right.  CO emission around 
-21 \kms\ is very widespread and vertically extended.  As noted by 
\cite{HewYus09} this velocity is inconsistent with normal rotation 
in the galactic disk but corresponds approximately to that of
the extended 3 kpc arm \citep{Ban80} which lies interior to the 
galactic molecular ring, within  the large-scale galactic bar 
\citep{Fux99}; the 3 kpc arm is not otherwise understood to be
so broad in latitude.  No especially 
strong CO emission from gas at this velocity is found anywhere
in Fig 1 (compare the upper left and lower right panels) but 
perhaps a yet-larger scale map would find clearer evidence of 
the SNR-neutral gas interaction in CO further to the South.
At a distance of 5 kpc the nominal \HH\ mass of the entirety
of the -21 \kms\ gas at l $<$ 5.4\degr, b $<$ -0.6\degr\ is 
$1.2\times10^4$ \Msun, 

\subsection{G5.55+0.32}

No maser spot was detected around this source by \cite{HewYus09} 
but strong (14K) CO emission is found at G5.64+0.23 (Fig. 2)
near the periphery of the remnant, along with a $\pm$ 9 \kms\ outflow 
which is marginally spatially resolved at our 1\arcmin\ resolution.  
As seen in the lower right panel of Fig. 1  there is 
an MSX source coincident with the CO peak and outflow,
and it is especially prominent in its field in MSX Band E 
at 22$\mu$ (not shown here).  

Is there a shell of neutral gas around the SNR G5.55+0.32? Perhaps a 
better question is where a candidate  shell is not found in Fig. 1.  
However, to the galactic south and southwest there is a very bright 
(peak 15 K) CO rim at v= 10-15 \kms\ just outside and seemingly tangent to the 
cartoon outline of the SNR (which is exceedingly faint and diffuse 
in the radiocontinuum).  Although in only in the very brightest or 
most kinematically or chemically disturbed gas will it be possible to 
establish an association in such a complicated region, 
such bright CO is seldom if ever seen in quiescent ambient dense gas 
in the general ISM.  At a distance of 2 kpc the nominal mass of \HH\ 
in the bright rim SW of the SNR is $3\times10^3$ \Msun; within a circle 
of radius 0.2\degr\ inscribed about the SNR the nominal mass 
is $1\times10^4$ \Msun\

\subsection{G5.71-0.08}

This SNR \citep{BroGel+06} coincides with $\gamma$-ray
source HESSJ1800-240C and molecular emission at 0-20 \kms\
near and just East of it (at l > 5.8\degr) was shown with much 
lower resolution NANTEN CO data (2.7\arcmin\ beam sampled 
at 4\arcmin\ intervals) by \cite{AhaAkh+08}.  

The peak CO temperature map at lower right in Fig. 1 shows that there 
is a rim of  bright CO emission around the SNR G5.71-0.08,  
with a rather narrow and very bright (19 K) CO line (Fig. 2) at the 12 
\kms\ 1720 MHz OH maser velocity, immediately adjacent to the maser 
spot.  Gas in the rim to the east of the remnant shows strong
positional velocity gradients, with more positive velocity (and
velocities above 15 \kms) to the north and west, which accounts
for the fact that a shell is more visible in the peak temperature
map.

All of the gas at 10-15 \kms\ southeast of G5.71 is bright,
as outlined in the peak temperature map at lower right in Fig. 1.
To the southeast, just  outside the bright rim there, is a very 
bright compact peak centered at G5.88,-0.13 and it is this gas
which appears in the NANTEN CO map shown by \cite{AhaAkh+08}.
The  bright CO in the western rim of the shell near the 1720 OH maser 
in our data is not apparent in the representation of the NANTEN data shown
by \cite{AhaAkh+08}.

A careful search of the CO profiles in this region did not show clear 
instances of the sort of broad-lined emission which sometimes characterizes 
shocked molecular gas, for instance in IC443 \citep{Den79}, at least
in part because of seeming confusion with other kinematic components;
the broader weaker emission at 0-30 \kms\ around the bright narrow
12 \kms\ feature at G5.704-0.12 in Fig. 2 is broadly-distributed
on the sky. \cite{Den83}  noted the 
difficulty of finding such evidence in a region as kinematically 
complicated as the galactic plane around l=6\degr, but 
shocked molecular gas near W28 was later found in more
complete mapping at higher resolution by \cite{AriTat+99}.

As shown by the v= -21 \kms\ CO near the OH maser spot around the 
SNR G5.4-1.2, CO lines are by no means always bright in gas around 
SNR, even when an interaction is occurring \citep{AriTat+99}.  However, 
bright CO seems to be the rule in W28 and in the SNR nearest to it in 
the region surveyed here.  The presence of such
bright CO lines is always the result of some special interaction, but
whether that interaction is with the SNR or one of the H II regions
is a matter for further discussion.

At a distance of 2 kpc the nominal \HH\ mass in the ridge of bright 
12\kms\ emission containing the 1720 OH maser spot is 
$2\times10^3$ \Msun\ and a circle inscribed about the SNR which includes this ridge and
the gas west of the SNR encloses $4\times10^3$ \Msun.  Including
all of the bright gas to the east increases the mass to $7\times10^3$ 
\Msun .


\subsection{G5.89-0.39 and other thermal sources}

\begin{figure*}
\psfig{figure=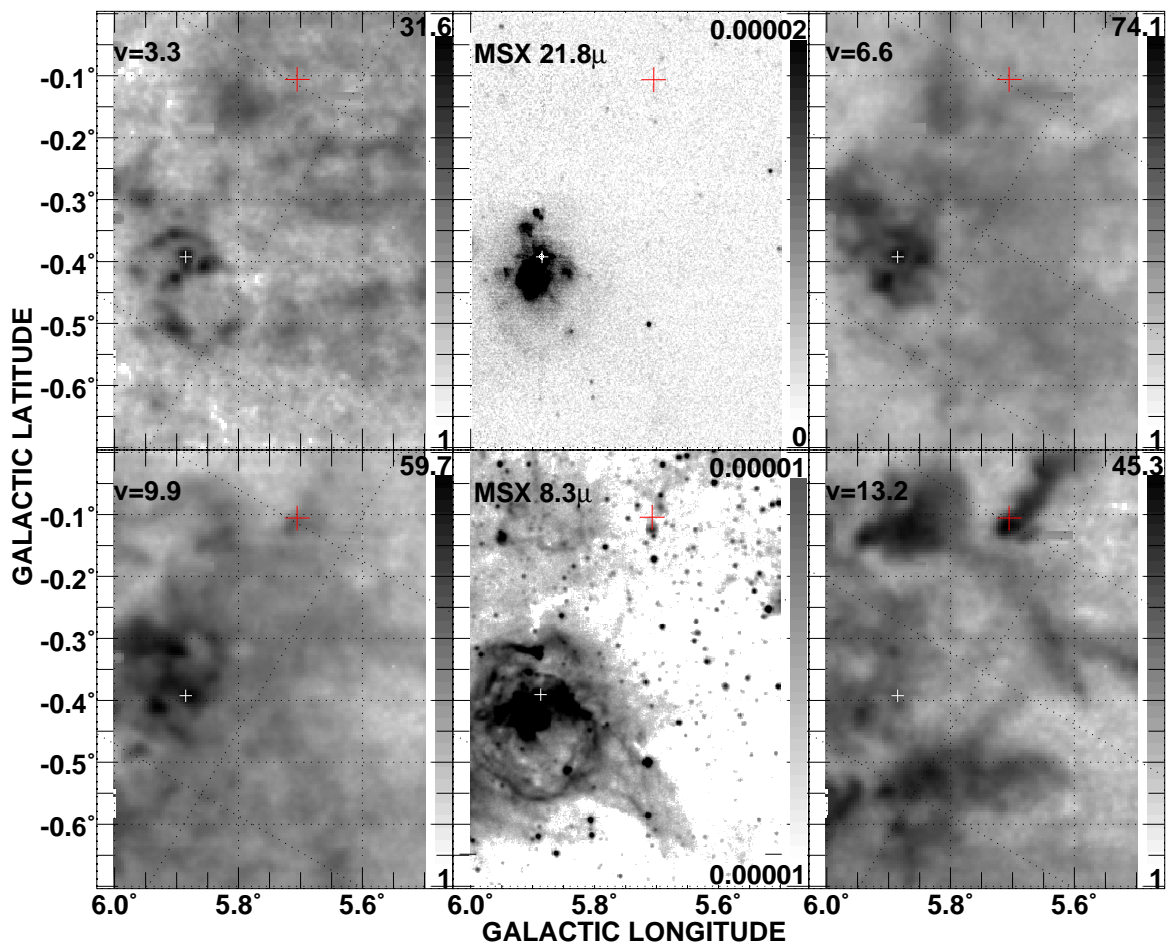,height=12.5cm}
\caption[]{ARO CO and MSX IR maps of W28A2 (G5.89-0.39) and its 
surroundings.  The middle panels show MSX Band A (bottom) and E (top) IR
emission, while the outer panels show integrated brightness of CO J=1-0
over 3-channel wide (3.047 \kms) intervals centered at the indicated velocities.
The position of the 1720 MHz OH maser discovered by \cite{HewYus09} is marked
by the upper (red) cross.}
\end{figure*}

Figure 3 shows in greater detail the distribution of CO and IR emission 
in the vicinity of the HII region W28A2 = G5.89-0.39. The well-studied 
bipolar outflow is visible in our data at the marked location but 
unresolved.  Neglecting the outflow, CO emission associated with the 
body of W28A2 occurs  mostly at v $\la$ 10 \kms\ and the strong 
emission at b =-0.125 \degr\ just north of W28 in the panel of Fig. 3
centered at 13 \kms\ is actually a distinct kinematic system which 
probably represents a separate H II region near the +12 \kms\
velocity of the 1720 MHz OH maser spot.

The nominal \HH\ mass of the molecular gas overlaying the body
of W28A2 and interior to the pseudo-shell shown in the lower
right panel of Fig. 3 is $2\times10^4$ \Msun.

\section{Summary}

We discussed CO J=1-0 emission observed by the ARO Kitt Peak 12m telescope
with 1\arcmin\ and 1\kms\ resolution  in the region 5\degr $< l < 6$\degr,
 -1.0\degr $< b < 0.5$\degr, just west of the SNR W28.  The mapped region contains
the compact HII region W28A2 (G5.89-0.39) and other more diffuse thermal ionized
gas (as seen at mid-IR wavelengths in MSX data), several candidate SNR (one
of which, G5.71-0.08 is coincident with the $\gamma$-ray source HESS J1800-240C)
and several newly-discovered 1720 MHz OH maser spots signifying interaction between 
SNR and ambient molecular gas.  The region is also host to a wide variety of
behavior in the CO emission, including a newly-discoverd outflow source and very
bright CO emission associated with both the thermal and non-thermal compact sources.

We showed the CO J=1-0 emission at the velocities of three newly-discovered 
1720 MHz OH maser spots. At -21 \kms\ CO, associated with G5.4-1.2, CO emission 
is widespread and emission near the maser is weak (3 K) and mundane. 
By contrast, CO emission around the other two SNR in the region is stronger
and more remarkable.  There is a bright (14 K) newly-detected CO outflow
source near G5.55+0.32 (which lacks a maser) and a rim of strong (15 K) CO 
emission just outside the remnant. CO emission at the OH maser spot 
toward G5.71-0.08 is extremely strong (19 K) and arcs of bright CO emission 
are seen around most of the periphery of this SNR.   Typical masses of the 
molecular clouds around the SNR are 10$^4$\Msun.

A concentration of strong CO just east of G5.71-0.08, previously observed in 
a low-resolution NANTEN map on a 4\arcmin\ pixel grid, overlays diffuse thermal
(MSX) emission and an MSX point source at l=5.9\degr, b=-0.13\degr\ which is 
probably the exciting source of a separate, classical HII region seen just north
of the stronger thermal source W28A2, G5.89-0.39.  Molecular gas associated
with the compact outflow source in G5.89-0.39 has been widely discussed but the
larger-scale distribution of CO emission around this HII region was shown in
detail here for the first time.  It does not show any obvious sign of having
been influenced by interaction with the outflow.

The profusion of compact thermal and non-thermal sources, thermal ionized gas 
and molecular clouds in the region presents an opportunity for study of the 
interaction between various phases of stellar evolution and the ISM, and indeed
further study will be required to understand, as it were, just who is doing what 
to whom, and with what.

\begin{acknowledgements}

The National Radio Astronomy Observatory is operated by Associated 
Universites, Inc. under a cooperative agreement with the US National 
Science Foundation.
The Kitt Peak 12-m millimetre wave telescope is operated by the
Arizona Radio Observatory (ARO), Steward Observatory, University
of Arizona.  I am grateful to the ARO Director, Dr. Lucy Ziurys. 
for awarding the observing time necessary to perform these 
observations and to the ARO staff and 12m operators who keep 
the telescope running at such a laudably high level.

\end{acknowledgements}

\bibliographystyle{apj}

\end{document}